# scientific reports

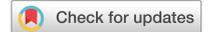

OPEN

# Fish evacuate smoothly respecting a social bubble

Renaud Larrieu[1], Philippe Moreau[1], Christian Graff[2], Philippe Peyla[1] & Aurélie Dupont[1]✉

Crowd movements are observed among different species and on different scales, from insects to mammals, as well as in non-cognitive systems, such as motile cells. When forced to escape through a narrow opening, most terrestrial animals behave like granular materials and clogging events decrease the efficiency of the evacuation. Here, we explore the evacuation behavior of macroscopic, aquatic agents, neon fish, and challenge their gregarious behavior by forcing the school through a constricted passage. Using a statistical analysis method developed for granular matter and applied to crowd evacuation, our results clearly show that, unlike crowds of people or herds of sheep, no clogging occurs at the bottleneck. The fish do not collide and wait for a minimum waiting time between two successive exits, while respecting a social distance. When the constriction becomes similar to or smaller than their social distance, the individual domains defined by this cognitive distance are deformed and fish density increases. We show that the current of escaping fish behaves like a set of deformable 2D-bubbles, their 2D domain, passing through a constriction. Schools of fish show that, by respecting social rules, a crowd of individuals can evacuate without clogging, even in an emergency situation.

Collective motion of animals is one of the most fascinating phenomena observed in nature: common examples include flocks of birds, herds of animals, colonies of ants and schools of fish[1,2]. This collective motion can be driven by a gregarious instinct to forage or mate or may be provoked by a panic situation. With humans, for instance, crowd movements can lead to stampedes, causing dramatic consequences, including fatalities. The worst-case scenario is the escape through a narrow passage in life-or-death circumstances. This situation has been extensively studied and simulated with the practical goal of improving escape efficiency in real-life situations[3,4]. With the help of simulations and an agent-based model, Helbing et al. described the faster-is-slower effect: the higher the desired velocity of the agents, the longer it takes to evacuate the room[5]. The evacuation becomes intermittent due to clogging events at the exit. The clogging of particles flowing through a bottleneck is actually relatively universal, as shown more recently by Zuriguel et al.[6], who identified the similarities between herds of sheep, pedestrian crowds, and even grains and colloids. By combining experiments and simulations, they proposed a unified framework for this clogging behavior, characterized by a power-law tail in the time lapse probability function[7]. So far, only ants have been observed to behave differently, managing a constant and optimized flow, even in high stress situations[8,9].

As in the seminal work of Helbing et al.[5], to test the "faster-is-slower" effect the usual control parameter is the desired velocity of animals, also called competitiveness[7] or the driving force for particles. However, the discharge of granular materials from silos has historically been studied with respect to another control parameter which is the ratio between the radius of the opening and the radius of the particles[10]. As expected, when the outlet size increases the clogging probability decreases, and the avalanche size increases non linearly[11], in addition, a critical radius exists above which no jamming appears[12]. This question has also been addressed for soft particles[13,14] where the link between the clogging probability and the deformability of particles has been evidenced[14]. Recently, Al Alam et al.[15] measured the phase diagram of the jamming probability of microalgae versus the door size and the swimming velocity in an experimental situation of evacuation through a bottleneck. This diagram confirms that decreasing the aperture size or increasing the agent's velocity result in the increase of the jamming probability, the transition from flowing to jamming exists along the two axes.

The typical evacuation experiment consists in forcing a group of animals to pass through a bottleneck, whose size is similar to the size of the animal. To the extent of our knowledge, this type of experiment has only been conducted on animals walking on a 2D solid substrate, except for the work by Al Alam et al. on microalgae[15]. However, this situation can also occur in liquids with schools of fish passing between rocks in rivers or on the seabed. Most fish exhibit gregarious behavior involving social interactions between individuals and traveling in groups[16–18]. Do jamming situations arise for fish? Intuition suggests that clogging is less likely because there is no solid friction between the fish. However, cognitive interactions could also lead to blockage situations and inefficient behavior. So far, the social interactions of fish have been studied in empty environments, free of any obstacles[19,20]. Here, we address this open question with a model of small fish in shallow water passing through various sized circular openings. We show that the evacuation of the fish follows a single statistical law of simple queuing, even though total evacuation takes longer for smaller apertures. Interestingly, for short time lapses between two fish exits, our results suggest a non-markovian process of short-time inhibition. The fish flow is suitably described by the same empirical law as for the evacuation of frictionless deformable objects in 2D. We propose to model each fish domain with a 2D bubble-like domain, whose size is governed by a *cognitive length*, i.e. the preferred distance between the fish, also called inter-individual distance in ethology[21].

[1]University Grenoble Alpes, CNRS, LIPhy, F-38000 Grenoble, France. [2]University Grenoble Alpes, CNRS, LPNC, F-38000 Grenoble, France. ✉email: aurelie.dupont@univ-grenoble-alpes.fr

## Methods

For our fish model, we chose the neon tetra *Paracheirodon innesi*, a small freshwater fish species originating from the western and northern Amazon basin (Fig. 1a). Being popular ornamental fish, *P. innesi* were easily found in aquarium stores supplied by fish farms. *P. innesi* exhibits robust shoaling behavior[22], and individuals rarely venture alone away from the group. We observed the smallest shoaling unit to be about 6 to 8 fish. *P. innesi* is a burst-and-coast swimmer, intermittently alternating active and passive phases. We can estimate their velocity to a few cm/s but this velocity is ill-defined. Thanks to the small size of the fish, about 3 cm long and 0.5 cm wide (Fig. 1a), the evacuation experiments could be set up easily on the bench in a standard fish tank (40x20 cm), divided in the middle by a home-made wall (Fig. 1b). A circular aperture, whose size could be modified from 4 cm to 1.5 cm using adapted ring-shaped covers, was made in the wall. For a typical evacuation experiment, a group of 30 fish was put into water about 10 cm deep in the experimental aquarium to obtain a quasi-2D system. After a rest and adaptation period of about 20 min, several passages of the group through the opening were recorded, with the fish being gently pushed towards the aperture with a fishing net. Images were taken from above using near infra-red lighting to avoid generating any additional stress for the fish, which are photophobic. The total duration of the fish experiments was limited to 1hr 30min per day for ethical reasons, and the water temperature was controlled (26 °C). In total about 80 different fish were used in the experiments, four fish died during this period in the housing aquarium. No fish were euthanized, all the fish were put back to the housing aquarium. The images were obtained using a Basler camera with a frame rate of 10 frames/s and analyzed with homemade codes in Python.

**Ethics.**   Our experimental protocols were approved by the ethics committee of the University Grenoble Alpes. Consisting of behavioral observations, they remained below the mild category of the severity classification of procedures, as defined by Section I of the Directive 2010/63/EU of the European Parliament and of the Council on the protection of animals used for scientific purposes. We followed the ASAB guidelines[25] and ARRIVE guidelines (https://arriveguidelines.org)".

## Results

**A cognitive length.**   The first step of data analysis consisted in manually counting each fish exit over time. Figure 1c shows the cumulative fish counts for three examples of evacuation through three different openings: diameter 4 cm, 2 cm and 1.75 cm. As we expected, the larger the opening, the faster the evacuation. Several experiments were averaged for each diameter ($11 \leq N \leq 17$) to obtain robust results; these are shown on Figure 1d. On average, for each diameter, the evacuation process occurred at a constant rate, except for the few last fish. The evacuation current, in fish per second, was obtained from linear fits of every realizations at a given diameter (excluding the 5 last fish) and then averaged to obtain the mean fish current and the standard error of the mean for all diameters. As shown on Fig. 2a, the evacuation current (fish/s) increases from 1 fish/s to 3 fish/s for opening diameters varying from 1.5 cm to 4 cm. From the experimental images, we also measured the fish density just before the opening by counting the number of fish present in a virtual 2D box of $8\times5cm^2$ (Fig. 1b, red dashed box). During the first half of the evacuation, fish density fluctuates around an average value for a given diameter (Fig. 2b). During the second part of the evacuation, the virtual box is emptied as the last fish evacuate, hence density decreases. The average density during evacuation of the first 15 fish is plotted for each diameter in Fig. 2c. Fish density remains stable for diameters larger than 2 cm at $\overline{\sigma} = 0.21 \pm 0.01 \ cm^{-2}$. For diameters smaller than 2 cm, fish density increases, denoting an accumulation of fish in the waiting zone in front of the aperture. However, the density for the smallest opening, $\sigma = 0.30 \pm 0.06 \ cm^{-2}$, is still far from the theoretical maximal density of fish in close contact, $\sigma_{max}^{th} \simeq 0.67 \ cm^{-2}$. Fish density seems to be controlled by social interactions for diameters of more than 2 cm. Above this specific density value, we defined $L_C = 1/\sqrt{\overline{\sigma}} = 2.18 \ cm \pm 0.05 \ cm$, the characteristic distance between the centers of two individual fish waiting to exit. Furthermore, we examined the number of fish per exit in detail. Given the temporal resolution ($dt = 0.1s$) for large openings, several fish were able to exit between two successive images (Supplementary Fig. S1). Considering only frames with exit events, the average number of fish exits per frame was 1.5 for the largest diameter ($D = 4 \ cm$) and this value decreased as the opening diameter got smaller, until the limit value of one fish per exit was attained in a dropper regime for diameters smaller than $L_C$. This regime coincided with the increase in fish density in the waiting zone, indicating that, for these diameters, the fish could not evacuate at the desired speed. For openings larger than $L_C$, the fish seem to respect social rules, waiting to exit at a specific density $\overline{\sigma}$. When the opening was smaller than the cognitive length, this behavior was frustrated: the fish started to accumulate at the exit. The fish current decreased to openings of about half the cognitive length, below which evacuation is impossible. It is interesting to note that there was no physical contact or friction limitation for the fish to pass through circular openings larger than 1.5 cm.

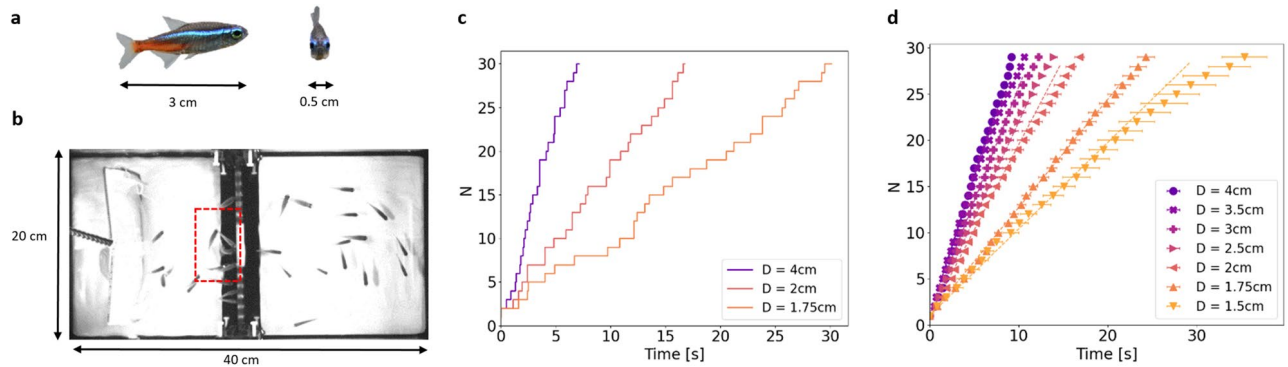

**Figure 1.**   (**a**) The biological model is *Paracheirodon innesi*, a small and common freshwater species. (**b**) Photograph of the experimental setup as imaged for data analysis. The tank is divided into two equal spaces by a wall with a circular opening of different sizes. The fish are forced to move toward the door by a net. The red dashed box is the area where the fish density is measured. (**c**) Number of evacuated fish as a function of time for three realizations with different opening diameters. (**d**) Averaged passage of 30 fish for different diameters (15 realizations for each diameter).

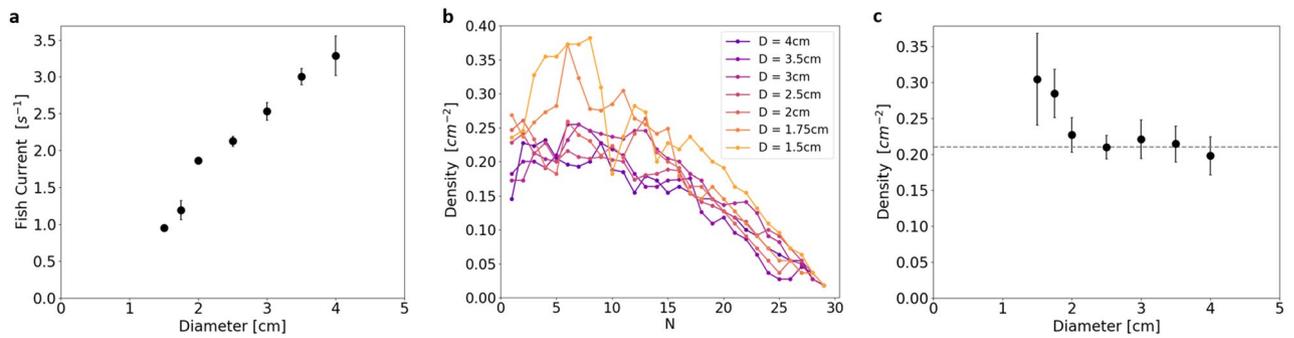

**Figure 2.** (**a**) Fish current, determined by the fit of every evacuation trials, as a function of the opening diameter. (**b**) Average fish density in a virtual box ($8\times5cm^2$, drawn on Fig. 1b) just before the opening as a function of the number of evacuated fish for the different opening diameters. (**c**) Average fish density over the first 15 fish as a function of the opening diameter. The density is stable for all diameters larger than 2 *cm*. A critical length $L_C$ is calculated from this averaged fish density (dashed grey line).

**A single statistical law.** Following the same approach as Zuriguel et al.[6], we calculated the time lapses between two fish exits and computed the complementary cumulative distribution function (CDF), i.e. the distribution of time lapses larger than $\tau$ (see Fig. 3a):

$$q(\tau) = \int_{\tau}^{+\infty} \rho(\tau')d\tau', \quad (1)$$

where $\rho(\tau)$ is the probability density function of the time lapses. The CDF corresponding to the two smallest diameters are distinctly separated from those with longer time lapses. This agrees with the previous results showing apparently distinct evacuation behavior for opening diameters smaller than the cognitive length $L_C$. Within the statistically relevant range and after a small delay, all the distributions seem to exhibit exponential decay (see Fig. 3b). Assuming that single fish exits are independent events and that the probability of an event occurring during $\delta\tau$ is $\delta\tau/\tau_0$, we can calculate that CDF exhibits exponential decay of characteristic time $\tau_0$ (Supplementary Information). This corresponds to standard queuing statistics. At short timescales, shouldering is observed, meaning that short time lapses are unlikely: each fish delays its exit after the preceding event.

Fish exits are not independent events at these short timescales. To take into account this delay at very short time scales and fit the distributions with a single law from $\tau = 0$, we created a function *ad hoc* based on the two limits: inhibition at a short time scale followed by exponential decay. Following the same approach as for demonstrating the exponential law, we propose:

$$\frac{1}{q}\frac{\delta q}{\delta \tau} = -\frac{1 - exp(-\tau/\tau_0)}{\tau_0}, \quad (2)$$

where $q$ is the probability that there is no exit at time $\tau$. For $\tau \gg \tau_0$, we once again observe the case of independent events leading to exponential decay of characteristic time $\tau_0$:

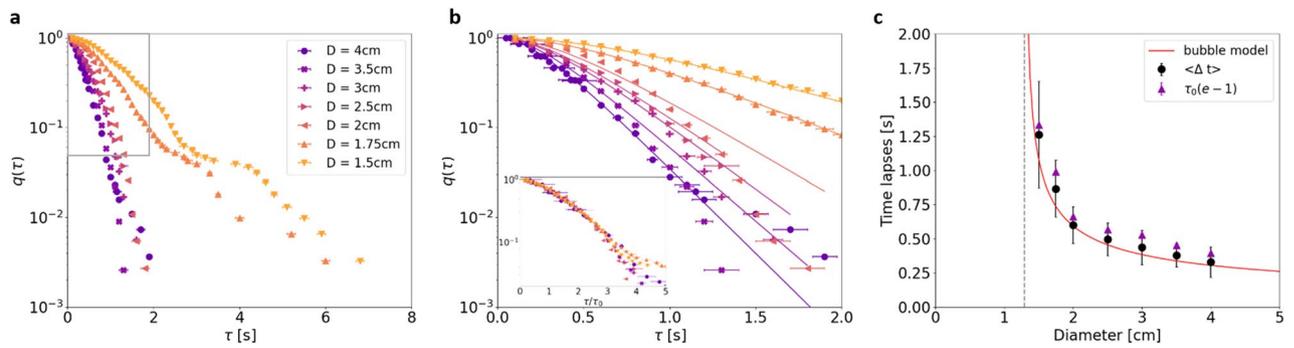

**Figure 3.** Statistical analysis of the fish evacuation. (**a**) Complementary of the cumulative distribution function of the time lapses between two fish exits, $q(\tau)$ for several opening diameters $D$, logarithmic scale on the $y$ axis. (**b**) Zoom in the previous plot overlaid with the corresponding single-parameter fits for each diameter. The fitting function is an interpolation between a short-time inhibition at short timescales and an exponential decay with characteristic time $\tau_0$ at longer timescales. In the inset are plotted the same data with respect to the rescaled time $\tau_0$ to show that the proposed model fits well the data for the whole range of diameters. **c**. The measured averaged time lapses for each diameter, $<\Delta t>$ are plotted together with the averaged time lapses obtained from the fit, $\tau_0(e-1)$. The two are in good agreement thereby validating the model. The dashed grey line shows $D_{stop}$, the diameter at which the flow vanishes and the time lapses diverge. The red line corresponds to Eq. (8) with the parameters obtained from the fit of the fish current.

$$\frac{1}{q}\frac{\delta q}{\delta \tau} \simeq -\frac{1}{\tau_0}. \tag{3}$$

But, when $\tau \ll \tau_0$, the approximation becomes:

$$\frac{1}{q}\frac{\delta q}{\delta \tau} \simeq \frac{\tau}{\tau_0^2} \tag{4}$$

and tends towards zero as $\tau$ tends towards zero. Other interpolations are possible when the limit at short timescales is zero and the behavior at longer timescales is exponential. After integration of Eq. (2), we obtain the statistical law used for single-parameter fitting in Fig. 3b:

$$q(\tau) = exp\left[1 - \frac{\tau}{\tau_0} - exp\left(-\frac{\tau}{\tau_0}\right)\right]. \tag{5}$$

The fits were obtained in the range where $q(\tau)$ is larger than $8 \cdot 10^{-2}$. With a total of about 360 experimental points, we estimated that uncertainty was too high below this probability. The clear collapse of the curves when plotted as a function of $\tau/\tau_0$ can be seen in the inset of Fig. 3b. The model therefore provides a good explanation of behavior for the range of diameters tested in this study. For the two smallest apertures, below the critical length $L_C$, egress is slower but follows the above-mentioned law without the characteristic power-law tail seen in clogging situations (see the log-log plot in Supplementary Fig. S2). Nevertheless, shouldering of these two distributions is visible in the log-lin plot Fig. 3a for time lapses of about 3 seconds. Such a shouldering has been observed by Pastor et al. in the case of granular materials in a vibrated hopper[7] and was interpreted as the crossover between a flowing regime at very short timescales and a clogging regime at longer timescales. Concerning the evacuation of fish, the shouldering is seen only on a couple of points that represent too few events to be able to reach a valid conclusion. Nevertheless, we cannot totally exclude a different regime for diameters smaller than 2cm at these long timescales. This would be interesting to be further investigated with a larger number of fish per experiment and a larger number of openings ranging between 1 and 2 cm.

Interestingly, a single characteristic time $\tau_0$ seems to be enough to describe both short-time inhibition and exponential decay. This characteristic time can be compared with the averaged time lapses $<\Delta t>$ directly obtained from the evacuation data. From Eq. (5) and the characteristic times $\tau_0$ obtained from the fits (Fig. 3b), the average interval time can be calculated:

$$<\Delta t_{fit}> = \tau_0 \cdot (e - 1). \tag{6}$$

Except for a small systematic bias of less than 0.1s, the interval times obtained directly from the data and from the fit process follow exactly the same line (see Fig. 3c), thus validating the merits of the proposed statistical law.

### Model: fish domains as cognitive bubbles
We have seen that a characteristic length can be defined based on the preferred fish-to-fish distance in the waiting zone. This cognitive length is very close to the crossover opening diameter that differentiates a dropper situation, in which the fish accumulate and evacuate more slowly, from a wise queuing situation, in which fish density is maintained at a stable value. We have also shown that a single-parameter statistical law can fit the survival function of the time lapses over all the opening diameters tested in this work. This law includes an inhibition at short time lapses which seems important for diameters smaller than the cognitive length. This can be interpreted as a difficulty to pass through a small aperture even though the aperture is still about twice the size that would physically constrain the fish. We propose to imagine the fish domain as a deformable 2D bubble whose diameter equals the cognitive length $L_C$.

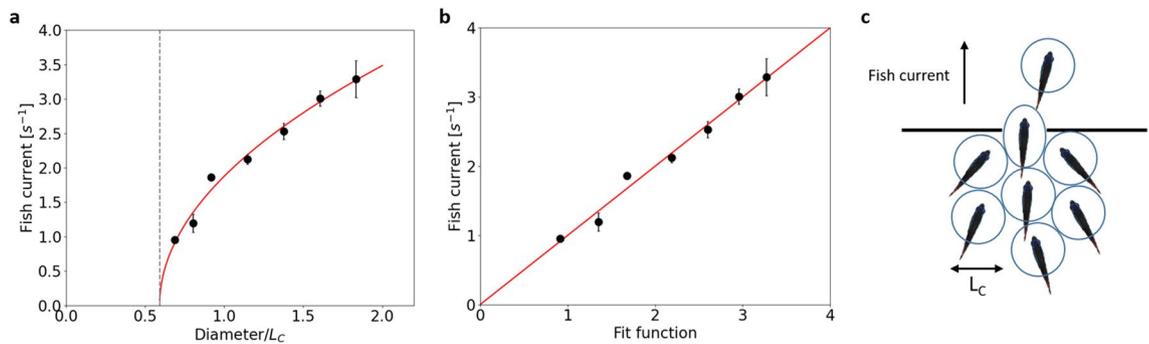

**Figure 4.** Fish as cognitive bubbles. (**a**) The modified Beverloo's law fits well the fish current when the cognitive length $L_C$ is taken as the characteristic size of the deformable particles. (**b**) Fish current plotted versus the fitted function to evaluate the goodness of the fit. The red line is $y = x$ as a reference. (**c**) Schematics of the model of cognitive bubbles dictating the fish behaviour.

Bertho et al.[23] have measured the flow of bubbles through an orifice where air bubbles were driven by Archimedes' forces in a 2D geometry. In this case, the bubbles are also pushed by their followers and energy can be dissipated in the deformation of the bubbles as well as in a viscous drag. Friction between the bubbles is negligible unlike the classical solid grains discharge. In this experimental work, both the driving force and the opening size were varied. Bertho et al. have shown that the bubble flow can be described by a modified Beverloo's law where the flow rate varies as $D^{1/2}$ unlike the classical Beverloo's law[10] for non-deformable solid grains where the exponent is 3/2. Here is the modified law that takes into account particle deformability:

$$j = A \left( \frac{D}{L_C} - k \right)^{1/2}, \qquad (7)$$

where $j$ is the particle flow rate, $D$ is aperture diameter, $L_C$ is bubble size, $k$ is related to particle deformability, and $A$ is a proportionality constant. $k$ is about 1.4 for granular solids that can cause clogging in apertures larger than the grain size due to the formation of arches. For bubbles, $k$ is about 2/3, as the bubbles can deform to pass through apertures that are smaller than the typical bubble size. Figure 4a shows the fit of the fish current to the Eq. (7) with two fitting parameters: $A = 2.94 \, s^{-1}$ and $k = 0.59$. "Bubble size" was set to the cognitive length $L_C = 2.18 \, cm$ and the exponent set to 0.5. To better evaluate the goodness of the fit, Fig. 4b displays the data points versus the fitting function. All the points are scattered across the $y = x$ line without a clear trend on one side or the other. If the exponent in Eq. (7) was let as a third free fitting parameter, the fitted exponent was 0.52, confirming the strong similarity with the law proposed for bubbles. In the fish case, the bubbles are virtual domains coordinated by cognitive rules. The driving force is more noisy than in the ideal case of identical air bubbles, the urge to evacuate can differ from fish to fish. Interestingly, the $k$ parameter is very close to that of air bubbles and is smaller than one, which is the signature of deformable particles.

According to Eq. (7), the model current is expected to vanish for an opening diameter: $D_{stop} = k \cdot L_C = 1.29 \, cm$. In practice, we were unable to perform any reproducible evacuation experiments for diameters below 1.5 cm, even though individual fish are able to swim through smaller apertures, down to about their body width, when attracted by food. The averaged time lapses $< \Delta t >$, as calculated previously, are also given by $< \Delta t_{bubble} > = 1/j$ and hence:

$$< \Delta t_{bubble} > = \frac{1}{A(\frac{D}{L_C} - k)^{1/2}}. \qquad (8)$$

According to this model, the average time lapses are expected to diverge for $D = D_{stop}$, which is indeed the case, as shown on Fig. 3c (dashed grey line). This figure displays as a function of the opening diameter the average interval times obtained by three different ways: (i) directly from the raw data $< \Delta t >$, (ii) from the single-parameter fitting of the CDF, $\tau_0 \cdot (e - 1)$, and (iii) from the bubble model $< \Delta t_{bubble} >$. The three methods are in very good agreement confirming the validity and consistency of both the fitting model of the statistical analysis and the bubble model for the fish current analysis.

To be even more precise in our analogy, we can consider the compact piling of circular bubbles that cannot achieve more than 91% occupancy. The measurement of cognitive length from fish density can be corrected for this geometrical factor and thus reaches $L_C = 2.35 \, cm$. Cognitive length was slightly underestimated by the 2D projection and analysis. This underestimation of cognitive length, which is a rather abstract length, led to a slight overestimation of the parameter $k$. In short, the cognitive bubbles of the fish are more deformable than the actual bubbles of Bertho et al.[23] We propose to model the short-time inhibition of one fish to follow another fish according to the concept of cognitive bubbles illustrated in Fig. 4c. This bubble is defined by the preferred fish-to-fish distance and must be deformed to pass through a small aperture. This behavior does not result in any clogging under the experimental conditions used in this work. Individual fish take longer to pass through the smallest apertures, but the queuing still follows an exponential rule with a characteristic waiting time.

## Conclusion

The flow of animals and hard particles through a bottleneck seems to be quite universal: except for ants, clogging occurs when the urge to evacuate is too high. This behavior is revealed by an intermittent flow leading to a power-law tail in the time lapse statistics. This question had not been previously addressed in liquids for cognitive swimming animals, nor in air for flying animals. Here, we tested the bottleneck evacuation on small gregarious fish in shallow water. As for ants, the egress statistics exhibit exponential decay, with fish patiently queuing, maintaining fish density in the waiting area. Even when the small size of the opening leads to an accumulation of fish, the time lapse statistics could still be fitted with a characteristic time. This fish accumulation starts when the opening becomes smaller than the preferred fish-to-fish distance. Based on this cognitive length, we propose an analogy with non-active matter. Indeed, the fish evacuation follows the same law as the passage of bubbles through a bottleneck. With the simple experiment described here, we obtained a model that provides a quantitative description of fish behavior and introduces a cognitive parameter into the modelling of active matter. Pedestrians and sheep tend to behave like solid grains in a vibrated hopper[7] where clogging occurs when the driving force is too large or the opening is too small. Fish tend to behave like deformable particles under a static load[23] where the clogging probability decreases as the driving force or the opening size are increased[14]. A recent study by Echevarria-Huarte et al.[24] may bridge the gap between the two scenarii: they show that when pedestrians are asked to evacuate while keeping a prescribed physical distance, no clogging appears and evacuation gets faster when the agent's velocity is higher. We show that, for this type of fish, social rules always appear to dominate, especially compared with sheep or humans, who tend to forget about social rules under high pressure. The "lessons" of this school of fish strengthened by the recent work by Echevarria-Huarte et al. could inspire swarm robotics, and help with traffic management for autonomous cars or human crowds. Ultimately, learning from fish behavior could help to improve the prevention of human stampedes, which repeatedly cause fatalities all around the world.

## Data availability

All data analysed during this study are included in this published article and its supplementary information files (figures of the raw data and complete analysis table). The raw dataset (movies and tracking data) used and analysed during the current study are available from the corresponding author on reasonable request.

## References
1. Couzin, I. D. & Krause, J. Self-organization and collective behavior in vertebrates. *Adv. Study of Behav.* **32**, 1–75. https://doi.org/10.1016/S0065-3454(03)01001-5 (2003).
2. Camazine, S. *et al. Self-Organization in Biological Systems* (Princeton University Press, UK, 2001).
3. Helbing, D. & Johansson, A. Pedestrian, crowd and evacuation dynamics. In *Encyclopedia of Complexity and Systems Science* (ed. Meyers, R. A.) 6476–6495 (Springer New York, New York, 2009). https://doi.org/10.1007/978-0-387-30440-3_382.
4. Zheng, X., Zhong, T. & Liu, M. Modeling crowd evacuation of a building based on seven methodological approaches. *Build. Environ.* **44**, 437–445. https://doi.org/10.1016/j.buildenv.2008.04.002 (2009).
5. Helbing, D., Farkas, I. & Vicsek, T. Simulating dynamical features of escape panic. *Nature* **407**, 487–490. https://doi.org/10.1038/35035023 (2000).
6. Zuriguel, I. *et al.* Clogging transition of many-particle systems flowing through bottlenecks. *Sci. Rep.* **4**, 7324. https://doi.org/10.1038/srep07324 (2014).
7. Pastor, J. M. *et al.* Experimental proof of faster-is-slower in systems of frictional particles flowing through constrictions. *Phys. Rev. E* **92**, 062817. https://doi.org/10.1103/PhysRevE.92.062817 (2015).
8. Boari, S., Josens, R. & Parisi, D. R. Efficient egress of escaping ants stressed with temperature. *PLoS ONE* **8**, e81082. https://doi.org/10.1371/journal.pone.0081082 (2013).
9. Wang, S., Lv, W. & Song, W. Behavior of ants escaping from a single-exit room. *PLoS ONE* **10**, e0131784. https://doi.org/10.1371/journal.pone.0131784 (2015).
10. Beverloo, W. A., Leniger, H. A. & van de Velde, J. The flow of granular solids through orifices. *Chem. Eng. Sci.* **15**, 260–269. https://doi.org/10.1016/0009-2509(61)85030-6 (1961).
11. Arévalo, R. & Zuriguel, I. Clogging of granular materials in silos: Effect of gravity and outlet size. *Soft Matter* **12**, 123–130. https://doi.org/10.1039/C5SM01599E (2016).
12. Zuriguel, I., Pugnaloni, L. A., Garcimartín, A. & Maza, D. Jamming during the discharge of grains from a silo described as a percolating transition. *Phys. Rev. E* **68**, 030301. https://doi.org/10.1103/PhysRevE.68.030301 (2003).
13. Ashour, A., Trittel, T., Börzsönyi, T. & Stannarius, R. Silo outflow of soft frictionless spheres. *Phys. Rev. Fluids* **2**, 123302. https://doi.org/10.1103/PhysRevFluids.2.123302 (2017).
14. Hong, X., Kohne, M., Morrell, M., Wang, H. & Weeks, E. R. Clogging of soft particles in two-dimensional hoppers. *Phys. Rev. E* **96**, 062605. https://doi.org/10.1103/PhysRevE.96.062605 (2017).
15. Al Alam, E. *et al.* Active jamming of microswimmers at a bottleneck constriction. *Phys. Rev. Fluids* **7**, L092301. https://doi.org/10.1103/PhysRevFluids.7.L092301 (2022).
16. Pavlov, D. S. & Kasumyan, A. O. Patterns and mechanisms of schooling behavior in fish: A review. *J. Ichthyol.* **40**, S163–S231 (2000).
17. Delcourt, J. & Poncin, P. Shoals and schools: Back to the heuristic definitions and quantitative references. *Rev. Fish Biol. Fisheries* **22**, 595–619. https://doi.org/10.1007/s11160-012-9260-z (2012).
18. Lopez, U., Gautrais, J., Couzin, I. D. & Theraulaz, G. From behavioural analyses to models of collective motion in fish schools. *Interface Focus* **2**, 693–707. https://doi.org/10.1098/rsfs.2012.0033 (2012).

19. Gautrais, J. *et al.* Deciphering interactions in moving animal groups. *PLoS Comput. Biol.* **8**, e1002678. https://doi.org/10.1371/journal.pcbi.1002678 (2012).
20. Katz, Y., Tunstrom, K., Ioannou, C. C., Huepe, C. & Couzin, I. D. Inferring the structure and dynamics of interactions in schooling fish. *Proc. Natl. Acad. Sci.* **108**, 18720–18725. https://doi.org/10.1073/pnas.1107583108 (2011).
21. Warburton, K. & Lazarus, J. Tendency-distance models of social cohesion in animal groups. *J. Theor. Biol.* **150**, 473–488. https://doi.org/10.1016/S0022-5193(05)80441-2 (1991).
22. Romano, D. & Stefanini, C. Individual neon tetras (Paracheirodon innesi, Myers) optimise their position in the group depending on external selective contexts: Lesson learned from a fish-robot hybrid school. *Biosys. Eng.* **204**, 170–180. https://doi.org/10.1016/j.biosystemseng.2021.01.021 (2021).
23. Bertho, Y., Becco, C. & Vandewalle, N. Dense bubble flow in a silo: An unusual flow of a dispersed medium. *Phys. Rev. E* **73**, 056309. https://doi.org/10.1103/PhysRevE.73.056309 (2006).
24. Echeverría-Huarte, I., Shi, Z., Garcimartín, A. & Zuriguel, I. Pedestrian bottleneck flow when keeping a prescribed physical distance. *Phys. Rev. E* **106**, 044302. https://doi.org/10.1103/PhysRevE.106.044302 (2022).
25. Guidelines for the treatment of animals in behavioural research and teaching. *Anim. Behav.* **83**, 301–309, https://doi.org/10.1016/j.anbehav.2011.10.031 (2012).

### Acknowledgements
The authors would like to thank Benjamin Dollet for fruitful discussions about the fish-bubble analogy, Catherine Quilliet for general discussions and mentoring of RL, Eric Bertin for theoretical discussions and Patrice Ballet for the first home-made fish tank used in preliminary studies. This work was funded by the Mission pour les initiatives transverses et interdisciplinaires (CNRS MITI call "Adaptation du vivant") and by the Institut Rhônalpin des systèmes complexes (IXXI). This research was funded in part, by the Agence Nationale de la Recherche (ANR PRC CES 45 FISHSIF). A CC-BY public copyright license has been applied by the authors to the present document and will be applied to all subsequent versions up to the Author Accepted Manuscript arising from this submission, in accordance with the grant's open access conditions.

### Author contributions
A.D., P.P. and C.G. designed the experiments. P.M. and R.L. set up the experimental system. R.L. performed the experiments. R.L. and A.D. did the analyses. A.D. drafted the manuscript. All authors reviewed the manuscript and approved the final version.

### Competing interests
The authors declare no competing interests.

### Additional information
**Supplementary Information** The online version contains supplementary material available at https://doi.org/10.1038/s41598-023-36869-9.

**Correspondence** and requests for materials should be addressed to A.D.

**Reprints and permissions information** is available at www.nature.com/reprints.

**Publisher's note** Springer Nature remains neutral with regard to jurisdictional claims in published maps and institutional affiliations.